\documentstyle[12pt,aps,prc,epsfig]{revtex}
\tightenlines

\begin{document}

\centerline{\bf Semiempirical Shell Model Masses with Magic Proton
Number} \centerline{\bf Z = 126 for Translead
Elements\footnote{Contribution to the 2nd Euroconference on Atomic
Physics at Accelerators: Mass Spectrometry, Carg\`ese, 19-23
September 2000.}}

\medskip

\centerline{\it S. LIRAN\footnote{Present address: Kashtan 3/3,
Haifa 34984, Israel}, A. MARINOV and N. ZELDES}

\centerline{\it The Racah Institute of Physics, The Hebrew
University of Jerusalem,}

\centerline{\it Jerusalem 91904, ISRAEL}

\begin{abstract}
A highly extrapolatable semiempirical shell model mass equation
applicable to translead elements up to Z = 126 is presented. The
equation is applied to the recently discovered superheavy nuclei
$^{293}118$ and $^{289}$114 and their decay products.
\end{abstract}

\section{Introduction}

A recent experiment \cite{nin99} is consistent with the formation
of the nucleus $^{293}$118 and its sequential decay down to
$^{269}$Sg (Z~=~106). The $\alpha$-decay energies vary rather
smoothly along the chain, precluding the traditional
macroscopic-microscopic \cite{mon94,smo97} Z~=~114 as a major
magic proton number in these nuclei. Recent phenomenological
studies of B(E2) \cite{zam95} and Wigner term \cite{zel98}
systematics indicate Z~=~126 as a plausible next spherical proton
magic number after lead. Recent self-consistent and relativistic
mean field calculations \cite{cwi96,ben99,kru00} variously predict
proton magicities for Z = 114, 120, 124 and 126, depending on the
interaction used.

Contrary to most of the above findings, the semiempirical shell
model mass equation (SSME) \cite{liz76} is based on the assumption
that Z = 114 is the next proton magic number after lead, and it
stops there. Moreover, the quality of its agreement with the data
starts deteriorating already beyond Hs (Z = 108). (See fig. 1, and
fig. 4 of ref. \cite{nin99}.) One has to find a substitute for the
equation in the neighbourhood of Z~=~114 and beyond.

In the early stages of developing the SSME \cite{liran} both
Z~=~114 and Z~=~126, then considered possible alternative
candidates for the postlead proton magic number, were tried as an
upper shell-boundary for translead elements. The agreement with
the data was about the same for both choices, and considering the
prevailing view in the mid nineteen-seventies Z = 114 was chosen
for the SSME mass table. We have recently \cite{lmz00} established
a high predictive power of the early Z = 126 results in the
interior of the shell region with Z~$\geq$~82 and 126 $\leq$ N
$\leq$ 184 (called here region B) by comparing them to the newer
data measured since then, and proposed using them as a substitute
for the SSME \cite{liz76} in superheavy elements (SHE) research.
We have also \cite{lmzsub} established a high predictive power for
$Q_{\alpha}$ values of the early Z~=~126 results in the interior
of the shell region with 82~$\leq$~Z, N~$\leq$~126 (called region
A). However, the quality of the predicted masses and other mass
differences worsened much compared to that of the original
adjustment. Therefore we readjusted the coefficients which largely
cancel in $Q_{\alpha}$, and proposed using the resulting equation
as a substitute for the SSME \cite{liz76} in the interior of
region A.

In sect. 2 we give the Z~=~126 equation in the two regions and
briefly discuss its predictive power. In sect. 3 we apply and
comment on it in relation to SHE research in region B.
%\vspace*{-5.0cm}
\begin{figure}[h]
\begin{center}
\leavevmode \epsfysize=5.0cm \epsfbox{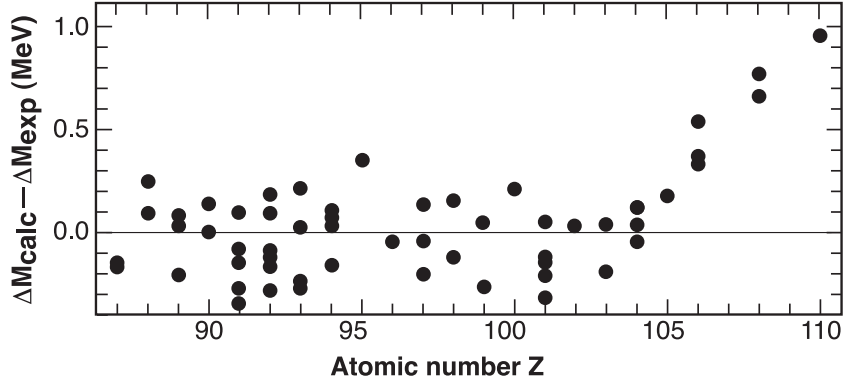}
\end{center} \caption{Deviations of the mass predictions
    [9] from the data for the 56 new masses in region B
    measured after the original adjustments were made.}
\end{figure}

\section{The Mass Equation and its Extrapolatability}

In the SSME the total nuclear energy E is written
\cite{liz76,zel96} as a sum of pairing, deformation and Coulomb
energies:

\begin{equation}
E\left( {N,Z} \right)=E_{pair}\left( {N,Z} \right)+E_{def}\left(
{N,Z} \right)+E_{Coul}\left( {N,Z} \right)\ , \label{eq1}
\end{equation}

\begin{equation}
E_{Coul}\left( {N,Z} \right)=\left( {{{2Z_0} \over A}}
\right)^{1/3} [{\alpha ^C+\beta ^C\left( {Z-Z_0} \right)+\gamma
^C\left( {Z-Z_0} \right)^2}]\  , \label{eq2}
\end{equation}

\begin{eqnarray}
E_{pair}\left( {N,Z} \right) &=& \left( {{{A_0} \over A}}
\right)[\alpha +\beta \left( {A-A_0} \right)+\gamma \left( {A-A_0}
\right)^2+\varepsilon T\left( {T+1} \right) \nonumber \\
  &&+{{1-\left( {-1} \right)^A} \over 2}\Theta +{{1-\left( {-1}
\right)^{NZ}} \over 2}\kappa ] \label{eq3}
\end{eqnarray}
for region A, and

\begin{eqnarray}
    E_{pair}(N,Z) &=& \left({{A_{0}}\over{A}}\right)[ \alpha +
\beta_{1} (N -
   N_{0}) + \beta_{2}(Z - Z_{0})  \quad  \nonumber \\
     &&+ \gamma_{1} (N - N_{0})^{2} + \gamma_{2}(Z - Z_{0})^{2}+
     \gamma_{3}(N-N_{0})(Z- Z_{0}) \quad \nonumber \\
     && +{{1 - (-1)^{N}}\over{2}}\Theta_{1} +{{1 -
     (-1)^{Z}}\over{2}}\Theta_{2} +{{1 - (-1)^{NZ}}\over{2}}\mu  ]
     \quad
    \label{eq4}
\end{eqnarray}
for region B,

\begin{equation}
E_{def}\left( {N,Z} \right)=\left( {{{A_0} \over A}} \right)\left[
{\varphi _{11}\Phi _{11}\left( {N,Z} \right)+\psi _{20}\left[
{\Psi _{20}\left( {N,Z} \right)+\Psi _{20}\left( {Z,N} \right)}
\right]} \right]
    \label{eq5}
\end{equation}
with
\begin{equation}
\Phi _{11}\left( {N,Z} \right)=\left( {N-82} \right)\left( {126-N}
\right)\left( {Z-82} \right)\left( {126-Z} \right) \ ,
    \label{eq6}
\end{equation}
\begin{equation}
\Psi _{20}\left( {N,Z} \right)=\left( {N-82} \right)^2\left(
{126-N} \right)^2\left( {N-104} \right)
    \label{eq7}
\end{equation}
for region A \cite{liran}, and

\begin{equation}
    E_{def}(N,Z)=\left({{A_{0}}\over{A}}\right)\left[\varphi_{21}\Phi_{21}(N,Z)+
\varphi_{31}\Phi_{31}(N,Z)+\chi_{12}X_{12}(N,Z)\right]
    \label{eq8}
\end{equation}
with
\begin{equation}
    \Phi_{21}(N,Z) = (N - 126)^{2}(184 - N)^{2}(Z - 82)(126 - Z) \ ,
    \label{eq9}
\end{equation}
\begin{equation}
    \Phi_{31}(N,Z) = (N - 126)^{3}(184 - N)^{3}(Z - 82)(126 - Z) \ ,
    \label{eq10}
\end{equation}
\begin{equation}
    X_{12}(N,Z) = (N - 126)(184 - N)(N - 155)(Z - 82)^{2}(126 -
    Z)^{2}(Z - 104)
    \label{eq11}
\end{equation}
for region B \cite{liran}.

In eqs.\ (\ref{eq2})$-$(\ref{eq5}) and (\ref{eq8}) $A=N+Z$ and $T
= |T_{z}| = {1 \over 2} |N - Z|$~\footnote{In the as yet unknown
odd-odd $N = Z$ translead nuclei the ground state (g.s.) is
expected to have $T = |T_{z}|+1$ and seniority zero, whereas eq.\
(\ref{eq1}) with $T = |T_{z}|$ gives the energy of a low excited
seniority two state \cite{zel96}}. The respective values of
($N_{0}, Z_{0}, A_{0}$) in regions A and B are (82, 82, 164) and
(126, 82, 208). The coefficients multiplying the functions of N
and Z are adjustable parameters determined by least squares
adjustment to the data, separately for region B \cite{liran} and
for region A \cite{liran,lmzsub}. Their  values are given in the
table.

We discuss extrapolatability first for region B. The experimental
data used in the adjustment \cite{liran} included 211 masses. Fig.
2 shows the deviations from the data of the predictions of eq.
(\ref{eq1}) for the 56 presently known new masses measured after
the adjustment \cite{liran}. The respective average and rms
deviations are 53 and 236 keV, as compared to 2 and 126 keV in
ref. \cite{liran}. The corresponding deviations of $Q_{\alpha}$
values are $-3$ and 220 keV, as compared to $-6$ and 162 keV.

The deviations of the seven N = 126-128 nuclei, denoted by empty
circles in the figure, increase when Z increases along the common
boundary of regions A and B away from the data. They are related
to the increasing discontinuity of the extrapolated mass surface
along the common boundary N = 126 of regions A and B away from the
data, when the two regions are adjusted separately
\cite{liz76,liran}. The deviations of the remaining 49 nuclei with
N $\geq$ 129, which do not follow the N = 126 boundary but extend
into the interior of the shell region, seem more random and they
are smaller, with respective average and rms deviations of $-1$
and 155 keV.

The above deviations are as a rule  about twice smaller than those
of several recent mass models. This is presumably due mainly to
the inclusion in eq. (\ref{eq1}) of the particle-hole $(p-h)$
symmetric
 configuration interaction terms $E_{def}$, eq.
 (\ref{eq8}). A more detailed discussion is given in
ref. \cite{lmz00}.

\begin{table}
    \caption{Values of the coefficients of
    eq.~(1) determined by adjustment to the data.}
    \begin{tabular}{crcccr}
    &&&&& \\
    %\hline \\
    \multicolumn{3}{c}{Region B \cite{liran}} &\multicolumn{3}{c}{Region
    A\cite{liran,lmzsub}} \\
        Coefficient & Value (keV) &&& Coefficient & Value (keV)  \\
        \hline
        $\alpha$ & $-2.3859605 \times 10^{6}$  &&&$\alpha$ &
$-1.987628 \times 10^{6}$\\
        $\beta_{1}$ & $-1.496441 \times 10^{4}$ &&&$\beta$ &
$-2.4773664 \times 10^{4}$ \\
        $\beta_{2}$ & $-3.3866255 \times 10^{4}$ &&&$\gamma$ &
$-8.51085 \times 10^{1}$\\
        $\gamma_{1}$ & $3.022233 \times 10^{1}$ &&&$\varepsilon$ &
$4.585496 \times 10^{2}$ \\
        $\gamma_{2}$ &  $2.811903 \times 10^{1}$ &&& $\Theta$ &
$1.2183 \times 10^{3}$\\
        $\gamma_{3}$ &  $-3.6159266 \times 10^{2}$ &&& $\kappa$ &
$2.1937 \times 10^{3}$\\
        $\Theta_{1}$ & $8.16 \times 10^{2}$ &&& $\alpha^{C}$ &
$7.968418 \times 10^{5}$ \\
        $\Theta_{2}$ & $1.007 \times 10^{3}$ &&& $\beta^{C}$ &
$2.032906 \times 10^{4}$ \\
        $\mu$ & $-2.121 \times 10^{2}$ &&&$\gamma^{C}$ & $9.819137
\times 10^{1}$ \\
        $\alpha^{C}$ & $8.111517 \times 10^{5}$ &&&$\varphi_{11}$ &
$-4.794 \times 10^{-2}$  \\
        $\beta^{C}$ & $2.0282913 \times 10^{4}$ &&& $\psi_{20}$ &
$9.095 \times 10^{-4}$ \\
        $\gamma^{C}$ & $1.0930065 \times 10^{2}$ &&&&  \\

        $\varphi_{21}$ & $-9.87874 \times 10^{-5}$ &&&&  \\

        $\varphi_{31}$ & $3.13824 \times 10^{-8}$ &&&& \\

        $\chi_{12}$ & $-1.428529 \times 10^{-7}$ &&&&  \\

    \end{tabular}
\end{table}

The situation in region A is less simple. The experimental data
used in the adjustment \cite{liran} included 29 masses and 62
$Q_{\alpha}$ values connecting unknown masses. The respective
average and rms predicted deviations for the presently known 121
new masses which became available after the adjustments increase
drastically to $-807$ and 1008 keV, as compared to $-29$ and 146
keV in ref. \cite{liran}. For the 31 new $Q_{\alpha}$ values,
though, the respective deviations are only 40 and 89 keV, as
compared to 5 and 103 keV.

\begin{figure}[h]
\begin{center}
\leavevmode \epsfysize=5.6cm \epsfbox{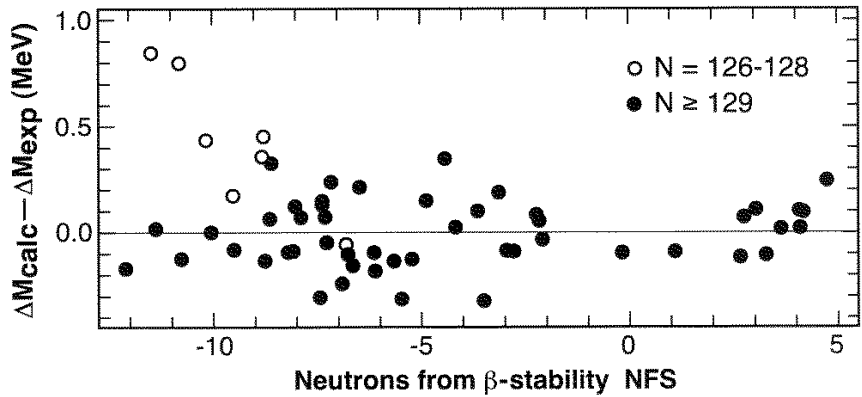}
\end{center} \caption{Deviations of the mass predictions
eq.~(1) from the data for
    the 56 new masses in region B measured after the original
adjustments were made.}
\end{figure}

In order to restore to the new mass predictions the same quality
as the old predictions had, while at the same time retaining the
high quality of $Q_{\alpha}$ predictions, we made \cite{lmzsub} a
least-squares adjustment of eq. (\ref{eq1}) to all the 150 known
masses, with only four adjustable parameters $\alpha, \varepsilon,
\Theta$ and $\kappa$ (eq. (\ref{eq3})) which largely cancel in
$Q_{\alpha}$, while the other seven coefficients were held fixed
on their old values \cite{liran}. These are the values given in
the table.

The readjusted values of $\alpha, \Theta$ and $\kappa$ are higher
and that of $\varepsilon$ is lower than in ref. \cite{liran},
indicating smaller overall binding, smaller symmetry energy
coefficient, and increased pairing energies in proton-rich nuclei
away from stability.

Fig. 3 shows the deviations from the data of the predictions of
the readjusted eq. (\ref{eq1}) for all the 150 experimentally
known masses. The respective average and rms deviations are 2 and
246 keV. The corresponding deviations of the predicted
$Q_{\alpha}$ values are 2 and 99 keV.

A more detailed discussion is given in ref. \cite{lmzsub}.

\begin{figure}[h]
\begin{center}
\leavevmode \epsfysize=5.6cm \epsfbox{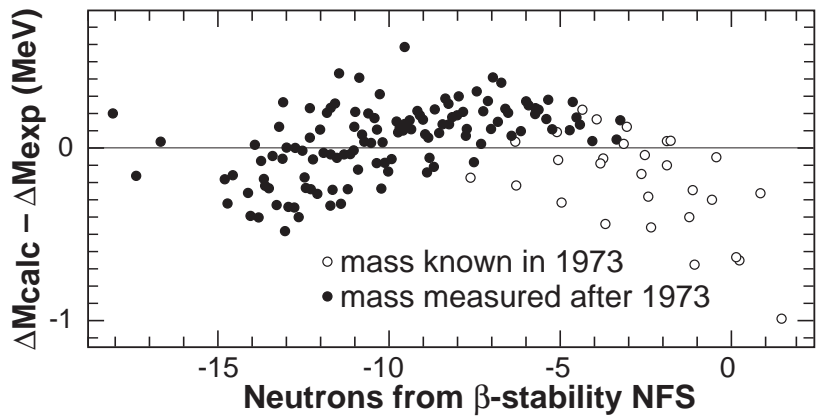}
\end{center} \caption{Deviations of the mass predictions
eq.~(1) from the data for
    all the 150 presently known masses in region A.}
\end{figure}

\section{Applications to SHE}

Panel a of fig. 4 shows the chain of measured $\alpha$-decay
energies starting from $^{293}$118 \cite{nin99}, and the values
predicted for them by eq. (\ref{eq1}) when the data are
interpreted as g.s. transitions of the assigned nuclei. The figure
shows as well the predictions \cite{smo99} which motivated the
search undertaken in \cite{nin99}. The respective average and rms
deviations of the predicted values from the data are $-197$ and
308 keV for eq. (\ref{eq1}) and $-154$ and 357 keV for ref.
\cite{smo99}. The rms deviation of eq. (\ref{eq1}) is consistent
with the deviations of the new nuclei in fig. 2 considered above,
but the average deviation is too negative. The respective average
and rms deviations of the predictions of eq. (\ref{eq1}) from
those of ref. \cite{smo99} are $-43$ and 369 keV.

The variation of the predicted values of eq. (\ref{eq1}) along the
chain is smoother than that of the data, where there are kinks at
Z~=~112 and 116 presumably representing submagic numbers or other
structure effects. In the SSME such effects are assumed
\cite{liz76,zel96} to have been smoothed out by configuration
interaction, represented by the terms $E_{def}$. The SSME is
inadequate to describe non-smoothed abrupt local changes
associated with subshell structure \cite{liz76}.

\begin{figure}[h]
\begin{center}
\leavevmode \epsfysize=12.0cm \epsfbox{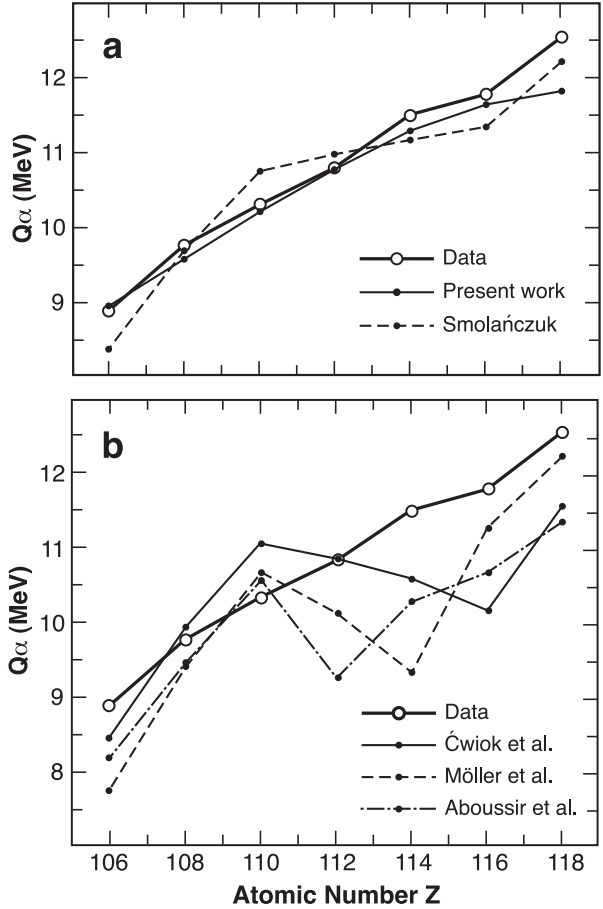}
\end{center} \caption{Experimental [1] and predicted
$Q_{\alpha}$
    values of the $^{293}$118 decay chain.
    (a) Predictions of eq.~(1) and of Smola\'nczuk [14].
    (b) Predictions of \'{C}wiok et al. [15], M\"oller et al.
        [16] and Aboussir et al. [17].}
\end{figure}

On the other hand, the microscopic energies calculated in ref.
\cite{smo99} are basically sums of (bunched minus unbunched)
single nucleon energies, and as such have (magic and) submagic gap
effects built in. The corresponding predicted line in fig. 4 has
kinks at Z = 110 and 116, corresponding to predicted submagic
numbers Z = 108 and 116.

Most of the smoothing effect of configuration interaction is
missing in macroscopic-microscopic Strutinsky type and in
self-consistent mean field calculations. The included $T=1, J=0$
pairing correlations seem not to be enough. This might result in
calculated submagic gaps and associated kinks which are too large
compared to the data. Panel b of fig. 4 shows such predicted large
kinks \cite{abo95,mnk97,cnh99}.

As a second application we address the $\alpha$-decay chain
observed in ref. \cite{oga99}, which is considered a good
candidate for originating from $^{289}$114 and its sequential
decay to $^{285}$112 and $^{281}$110. The respective average and
rms deviations of the predictions of eq.~(\ref{eq1}) from the
measured energies are 847 and 905 keV, which considerably exceed
the deviations expected from fig. 2 for g.s. transitions. If the
above assignments are confirmed, the large deviations might
indicate that the decay chain does not go through levels in the
vicinity of the g.s.

It might also be worthwhile mentioning, that for the conceivable
parents $^{288}$112 or $^{291}$113 which can be obtained from the
compound nucleus $^{292}$114 by respective $1 \alpha$ or $1p$
evaporation, the corresponding average and rms deviations of
eq.~(\ref{eq1}) from the measured energies are $-181$ and 366 keV
and $-242$ and 417 keV, which are more than twice smaller than for
the parent $^{289}$114.

In conclusion we address the bearing of the Z = 118 results on the
question of magicity of Z~=~126 versus 114. The energy eq.
(\ref{eq1}) comprises a part $E_{pair} + E_{Coul}$, eqs.
(\ref{eq2}) and (\ref{eq4}), which is a quadratic parity-dependent
function of the valence particle numbers, and a part $E_{def}$,
eq. (\ref{eq8}), which is a $p-h$ symmetric function of both
particle and hole numbers. For different assumed upper shell
boundaries the hole numbers are different and $E_{def}$ is a
different function of N and Z. The superior agreement of eq.
(\ref{eq1}) with the data in fig. 4, as compared to the
deterioration of the SSME \cite{liz76} near Z~=~114 mentioned in
the introduction, demonstrates the superiority of proton-hole
numbers defined with respect to Z~=~126 as a proton magic number
rather than Z~=~114. This conclusion has also been arrived at in
ref. \cite{zam95} on the basis of B(E2) systematics. It is
important to emphasize, though, that the above rather suggestive
results are not a proof of superior magicity of Z = 126 as
compared to the recently predicted Z = 120 or 124, because no
comparative mass studies of this kind were made.

We thank Stelian Gelberg and Dietmar Kolb for help with the
calculations.

\end{document}